\documentclass[a4paper,12pt]{article}
\usepackage{graphicx}
\usepackage[top=3.0cm,bottom=3.0cm,left=3.0cm,right=2.0cm]{geometry}
\usepackage{amssymb,amsmath}
\usepackage{textcomp}
\usepackage{subfigure}
\usepackage{feynmp}
\usepackage{cite}
\def\pacs#1{\vspace{10pt} \hspace{0.33cm} \rm PACS numbers: #1 \par \vspace{10pt}}

\title{Renormalization group equation for Tsallis statistics}
\author{Airton Deppman}
\date{1- Instituto de F\'isica -  Universidade de S\~ao Paulo \\ email: deppman@if.usp.br \\ Rua do Mat\~ao Travessa R Nr.187 CEP 05508-090 Cidade Universit\'aria, S\~ao Paulo - Brasil.}

\begin{document}

\maketitle

\begin{abstract}
The  non-extensive statistics proposed by C. Tsallis has found wide applicability, being present even in the description of experimental data from high energy collisions. A system with a fractal structure in its energy-momentum space, named thermofractal, was shown to be described thermodynamically by the non-extensive statistics. Due to the many common features between thermofractals and Hagedorn's fireballs, this system offers the possibility to investigate the origins of non extensivity in hadronic physics and in QCD. In this regard, the investigation of the scaling properties of thermofractals through the renormalization group equation, known as Callan-Symanzik equation, can be an interesting approach.
\end{abstract}


In the present work Tsallis statistics is analyzed in the context of renormalization theory. The relations between such  non-extensive statistics and scaling properties are expressed in terms of a Callan-Symanzik equation~\cite{Callan, Symanzik1, Symanzik2}, which represents the fundamental properties of a scale free system. 

The generalization of Boltzmann-Gibbs-Shannon (BGS) statistics by violation of entropy additivity mediated by the entropic index $q$ leads to Tsallis statistics~\cite{Tsallis1988}, which will lead to  non-extensive thermodynamical quantities that were expected to be extensive in the context of BGS. Tsallis statistics is  known to apply to a large number of systems in Physics and in other fields, and one of its most distinguished features is the power-law distribution in contrast to the exponential behavior common to BGS distributions. One of the most interesting applications of the generalized thermodynamics lies in the description of distributions found in high energy collisions experiments~\cite{Bediaga, Beck2000, Wong,WongWilk}. A generalized version of Hagedorn's self-consistent thermodynamics~\cite{Deppman2012} has allowed to predict a limiting temperature and a common entropic index, $q$, and a new hadron mass spectrum formula. The results found fair agreement with experiments~\cite{Sena, Cleymans_Worku, Lucas, Lucas2, AzmiCleymans2, DeBhaskar}.

The Callan-Symanzik equation was formulated in the context of renormalization theory of Quantum Gauge Fields with scale invariance. The Yang-Mills theory, in particular, is scale free and may satisfy that equation. In this regard, the Callan-Symanzik equation was fundamental to determine the asymptotic freedom of QCD~\cite{ Politzer1, Politzer2, GrossWilczek1, GrossWilczek2}.

In Refs.~\cite{Deppman2016, Deppman:2015cda, Deppman:2016prl} it was shown that a system with a particular fractal structure in the energy-momentum space should be described by the  non-extensive statistics proposed by Tsallis. Such system, named thermofractal, presents three fundamental properties:
\begin{enumerate}
 \item it has an internal structure formed by $N'$ thermofractals;
 \item the total energy of the thermofractal is the sum of the total kinetic energy, $F$, and the total internal energy, $E$, of the compound thermofractals. These energies are such that the ratio $E/F=\varepsilon/k\tau$ fluctuates according to the probability density $P(\varepsilon)$.
  \item The internal energy decreases as deeper levels of the thermofractals are considered.
\end{enumerate}

It is possible to show~\cite{Deppman2016} that the probability density for such system is given by
\begin{equation}
  P(\varepsilon)=\left[1+(q-1)\frac{\varepsilon}{k\tau}\right]^{-1/(q-1)}\,,  \label{qexp}
\end{equation}
where $\tau=(q-1)T$, with $T$ being the temperature of the thermofractal.
A consequence of such properties is that the temperature of thermofractals at level $n$ scales, on average, as
 \begin{equation}
 \frac{T^{(n)}}{T}= \frac{E^{(n)}}{E}\,. 
\end{equation}
This system can be shown to have a fractal dimension in the energy-momentum space, so from now on it will be referred to as fractal.
They are scale free systems and present several characteristics that are interesting to investigate the origin of non extensivity in hadron systems, as the similarities with Hagedorn's fireballs. With the introduction of this kind of fractal it was possible to understand that Hagedorn's theory, which is based on a self-referenced definition of fireballs or hadrons, should necessarily be described by Tsallis statistics. In the circumstances of hadron physics, it allows a new understanding on the intermittency effect~\cite{Bialas_Peschanski, Bialas_Peschanski2, Hwa, HwaPan, Sarkisyan, Sarkisyan2} observed in high energy data, determines the related fractal dimension and connects this effect to other features of high energy experimental data, such as self-similarity~\cite{WWselfsymmetry, Tokarev, Zborovsky}, long-tail distributions~\cite{WongWilk}, mass spectrum~\cite{Lucas}.

Intermittency effects, in particular, have been associated to fractal-like properties of the multiparticle production process~\cite{Bialas90, Peschanski,BraxPeschanski} (see~\cite{Kittel, Wolf} for a more complete account on the subject), and it was associated to gluon emission of high energy jets~\cite{Veneziano, Konishi, Konishi2} that results from the QCD evolution equations~\cite{AltarelliParisi}. These equations arises from the properties of the renormalization group for non-Abelian Yang-Mills gauge field theory~\cite{Politzer1, Politzer2, GrossWilczek1, GrossWilczek2}.

A detailed analysis of thermofractals and their properties allows one to show that the density in Eq.~(\ref{qexp}) can be written in terms of $F$ and $E$ for a fractal at an arbitrary level $n$ as
\begin{equation}
  P(F,E)= N'^n \left(\frac{F^{(n)}}{kT^{(n)}}\right)^{3/2} e^{-U^{(n)}/(kT^{(n)})} \left[1-(q-1)\frac{\varepsilon}{k\tau}\right]^{-1/(q-1)}\,, 
\end{equation}
with $U=E+F$. Introducing $M=kT$ for convenience and taking into account that
\begin{equation}
  \left(\frac{1}{N}\right)^{n/(1-D)}=\frac{T^{(n)}}{T}\,, 
\end{equation}
it results
\begin{equation}
  P(F,E)= \left(\frac{M^{(n)}}{M}\right)^{-(1-D)} \left(\frac{F^{(n)}}{M^{(n)}}\right)^{3/2} e^{-U^{(n)}/M^{(n)}} \left[1-(q-1)\frac{\varepsilon}{k\tau}\right]^{-1/(q-1)}\,, \label{nlevel}
\end{equation}
where $D$ is the Haussdorf fractal dimension~\cite{Deppman2016}.

Notice that for a fixed value of the scale $M$, hence at a fixed level $n$ of the fractal structure, the equation above is a well defined continuous function, and a simple analysis would lead one to conclude that the dimension $D$ is not fractal, but reflects the topology of the phase-space where the system is embedded. This is due to the fact that the anomalous dimension arises from the fractal structure itself, and not from the underlying distribution. In other words, it is necessary to take into account the fractal evolution with the scale variation, which leads to a tree-like diagram, to obtain the fractal dimension. A nice account on the subject, in general, can be found in Refs.~\cite{Sarkisyan, Sarkisyan3}, and for a specific description of the system analyzed here, in Ref.~\cite{DMMF}.

In the present work the scaling properties of the fractal structure will be investigated under the light of renormalization theory. In this sense the scaling properties can be analyzed in two ways: (a) by varying $E$ and $F$ while keeping $M$ fixed; (b) by varying $M$ while keeping $E$ and $F$ constant. Both transformations are equivalent according to scaling properties, and are related through the fundamental equation of renormalization theory, the Callan-Symanzik equation. The main objective here is to obtain such equation in the context of fractals. Before doing that, observe that since $E$ and $F$ are transformed by the same scale factor, the ration $E/F$ remains constant, and so remains the parameter $\varepsilon/(k\tau)$. In addition, the exponential factor in Eq.~(\ref{nlevel}) amounts to the Boltzmann factor for thermal equilibrium, and bears no relation to the fractal structure itself, so it must be  dropped for the analysis. With these considerations, the invariance of the fractal structure by scale transformation can be expressed by the identity
\begin{equation}
 \Gamma(F,M)= \left(\frac{M}{\Lambda}\right)^{-(1-D)} \left(\frac{F}{M}\right)^{3/2}\,, \label{renorm}
\end{equation}
with $\Lambda$ being some reference scale.

The equation above is suitable for the scaling analysis in both ways described above. From the method (a), where $M$ is fixed and $F$ varies, one gets
\begin{equation}
 F \frac{\partial \Gamma}{\partial F}= \frac{3}{2}\Gamma\,.
\end{equation}
From the method (b), where $F$ remains constant while the scale $M$ varies, one gets from Eq.~(\ref{renorm})
\begin{equation}
  M \frac{\partial \Gamma}{\partial M}= \left(-\frac{3}{2}-(1-D)\right)\Gamma\,.
\end{equation}
The results above allow one to obtain the Callan-Symanzik equation for the fractals considered here, i.e.,
\begin{equation}
\left[M \frac{\partial}{\partial M}+F \frac{\partial}{\partial F} + d\right]\Gamma=0\,, \label{CallanSymanzikTsallis}
\end{equation}
where $d=1-D$ is the anomalous fractal dimension.

The fractal dimension $D$ was determined in terms of the parameters that characterize thermofractals~\cite{Deppman2016}, and is given by
\begin{equation}
 D=1+\frac{\log N'}{\log R}\,,
\end{equation}
where
\begin{equation}
 R=\frac{(q-1)N/N'}{3-2q+(q-1)N}\,, \label{scale}
\end{equation}
with $N=N'+2/3$.

Eq.~(\ref{CallanSymanzikTsallis}) represents the fundamental properties of the fractal structure under scale transformation. Since it is related to a system which scaling properties are the main ingredient to obtain Tsallis statistics, one can recognize the equation above as the Callan-Symanzik equation for Tsallis statistics.

This result sets the ground for an interpretation of Tsallis statistics in association to renormalization group theory. In addition, it opens new possibilities to explore the potential applicability of  non-extensive statistics in the domain of hadronic physics, and hopefully allows for a deeper understanding of the properties of QCD that makes self similarity and fractal structures to emerge from the strong interaction in complex system. It can be also be associated to the non-thermal phase transition of hadronic matter associated to Quark-Gluon Plasma~\cite{Kittel, Peschanski2}.

In a thermodynamical approach, the application of the non-extensive self-consistent thermodynamics that arises from the fractal structure when applied to hadronic systems has gone already beyond the usual description of high energy distributions, and has been extended to systems with finite chemical potential~\cite{Megias, Megias:2014tha}, to extend the MIT Bag model by including a fractal structure~\cite{PedroCardoso} and to describe neutron star equilibrium~\cite{Debora}.

In conclusion, the Callan-Symanzik equation associated to Tsallis statistics was  derived here in association to the thermofractal scale free structure, setting new grounds for the interpretation of  non-extensive thermodynamics in terms of renormalization group theory, and opening new possibilities to its application in QCD related problems.

{\bf Ackowledgements} This work is supported by Conselho Nacional de Desenvolvimento Ciet\'ifico e Tecnol\'ogico, CNPq. The author declare that there is no conflict of interest regarding the publication of this paper.

\end{document}